# COMPUTING THE MINIMAL CREW
## for a multi-generational space journey towards Proxima Centauri b

**FRÉDÉRIC MARIN**[1] **& CAMILLE BELUFFI**[2]   [1] Université de Strasbourg, CNRS, Observatoire astronomique de Strasbourg, UMR 7550, F-67000 Strasbourg, France   [2] CASC4DE, Le Lodge, 20, Avenue du Neuhof, 67100 Strasbourg, France

**email**  frederic.marin@astro.unistra.fr

The survival of a genetically healthy multi-generational crew is of a prime concern when dealing with space travel. It has been shown that determining a realistic population size is tricky, as many parameters (such as infertility, inbreeding, sudden deaths, accidents or random events) come into play. To evaluate the impact of those parameters, Monte Carlo simulations are among the best methods since they allow testing of all possible scenarios and determine, by numerous iterations, which are the most likely. This is why we use the Monte Carlo code HERITAGE to estimate the minimal crew for a multi-generational space travel towards Proxima Centauri b. By allowing the crew to evolve under a list of adaptive social engineering principles (namely, yearly evaluations of the vessel population, offspring restrictions and breeding constraints), we show in this paper that it is possible to create and maintain a healthy population virtually indefinitely. An initial amount of 25 breeding pairs of settlers drives the mission towards extinction in 50 ± 15% of cases if we completely forbid inbreeding. Under the set of parameters described in this publication, we find that a minimum crew of 98 people is necessary to ensure a 100% success rate for a 6300-year space travel towards the closest telluric exoplanet known so far.

## 1    INTRODUCTION

In 2016, the closest-to-Earth exoplanet was discovered [1]. This exoplanet, Proxima Centauri b, is believed to be rocky with an inferred minimum mass of 1.27 +0.19/-0.17 Earth masses, which makes it a good target for a future manned exploratory mission. Proxima Centauri b is also slightly larger than Earth (0.94-1.40 $R_{Earth}$ [2]) and revolves with a period of 11.2 days on a stable, low-eccentricity orbit around our nearest stellar neighbour, Proxima Centauri. This red dwarf has a luminosity of about 0.15% that of the Sun but Proxima Centauri b is situated only at a semi-major axis distance of 0.05 astronomical units (AU) from the star. The exoplanet is thus irradiated by a stellar flux that is ~ 0.65 times that for Earth [3, 4], leading to an equilibrium temperature of 234 K [1]. Considering an atmosphere with a surface pressure of one bar, one could expect liquid water to be present on the surface of the planet [5, 6]. Proxima Centauri b is thus within the range of potential habitability and becomes an interesting target for an exploratory mission.

The distance towards Proxima Centauri is estimated to 1.295 parsecs [7]. This corresponds to about 4 ×$10^{13}$ km and it takes 4.22 years for light to reach us. The fastest human-made objects are far from reaching such high speeds and a manned mission to Proxima Centauri b would thus take much longer. As an example, the Apollo 11 spacecraft reached speeds near 40,000 km.$h^{-1}$, with an average velocity of about 5500 km.$h^{-1}$. Any space travel onboard of Apollo 11 would have taken approximately 114,080 years to reach Proxima Centauri b, disregarding food, water, oxygen or power supplies. If we aim at exploring or colonizing the closest exoplanet from the Earth, it is mandatory to find a faster spacecraft. One of the pioneering international attempts to design and study a space vehicle able to reach neighbouring stars was the Orion Project [8]. Based on nuclear energy, the Orion Project would have achieved both a strong thrust and a large specific impulse, in theory enabling very inexpensive large-scale space travel. Instead of a combustion chamber-nozzle configuration, Orion's concept was to eject fission charges up to a few hundred meters behind the vehicle, intercepting the plasma on a thick pusher plate made of steel or aluminium. Huge shock absorbers transformed the 50,000 g received by the plate into constant thrust. The theoretical cruising speed reached by a thermonuclear Orion vessel would have been 8-10% of the speed of light, but the project lost political support in 1963 because of concerns about propellant contamination [9]. At 10% of the speed of light, the travel would have taken 42 years – but to get it to interstellar velocities requires a truly stupendously large vehicle [10].

While the concept of external nuclear propulsion was abandoned, other concepts of spacecraft emerged such as the Daedalus project. This can be considered as the first really detailed study of an interstellar probe that gathered great minds between 1973 and 1978 [11]. More recently, the Breakthrough Starshot Initiative was started. It consists of a set of multiple, very light spacecraft that use lightweight sails to catch laser beams shot from Earth which accelerate them to one-fifth the speed of light [12]. Even if this project is appealing, the probes are not yet built and their essentially negligible mass confines such spacecraft to unmanned missions.

For a safer estimation of the human capabilities to reach high speeds, one should consider real missions that will fly in the coming years. By 2018, the NASA mission "Solar Probe Plus", recently renamed "Parker Solar Probe", will be launched. Its goals are to come as close as 8.5 solar radii to the Sun to trace the flow of energy that heats and accelerates the solar corona and solar wind; to determine the structure and dynamics of the plasma and magnetic fields at the sources of the solar wind; and to explore the mechanisms that accelerate and transport energetic particles [13]. The Parker Solar Probe will reach record breaking orbital velocities as high as 724,205 km.h$^{-1}$ (~ 200 km.s$^{-1}$), which translates into about 0.067% of the speed of light. At this speed, an interstellar journey would still take about 6300 years to reach Proxima Centauri b. While we can't yet actually launch ships at this speed, it shows that our technology is able to remain operational at high speeds. We will therefore use this 200 km.s$^{-1}$ velocity as a starting point for this paper. At 0.067% the speed of light, a futuristic interstellar journey cannot take less than one generation to reach the closest exoplanet to Earth. Future development in space engineering might reduce the time needed to reach Proxima Centauri b, but current estimates must rely on the assumption that multi-generational crews are needed to achieve interstellar exploration (other methods for space exploration are described in detail in [14]).

This is the aim of this paper: to quantify the minimal initial crew necessary for a multi-generational space journey to reach Proxima Centauri b with genetically healthy settlers. To do so, we use the Monte Carlo code HERITAGE presented in the first paper of the series [14] and run simulations accounting for a 6300-year trip at 200 km.s$^{-1}$. It is much longer in duration than what was investigated before, since we rely on the technology of the Parker Solar Probe instead of non-existent technologies; but this will allow us to determine if population stability over millennia is achieveable. In Section 2.1 we show that fixed rules within the ship are not a reasonable option for sustaining genetic diversity. We advocate adaptive social engineering principles and present in Section 2.2 the results of our computations. The impact of inbreeding restrictions on the population is studied in Section 2.3, and the rate of success for a variety of initial crew is presented in Section 2.4. We discuss the relevance of Proxima Centauri b as the first target to be explored by a multi-generational space expedition in Section 3 and conclude our paper in Section 4.

## 2 SIMULATING A JOURNEY TO PROXIMA CENTAURI b

To simulate a multi-generational space journey, we use the Monte Carlo code HERITAGE extensively presented in [14]. As a brief reminder, a Monte Carlo simulation is a computerized mathematical technique that takes into account chance events in decision making. The code accounts for a wide range of possible outcomes and their probabilities of occurrence depending on randomized actions. It reveals the extreme possibilities, as well as all the possible consequences of intermediate decisions. Applied to interstellar multi-generational spacecraft, such a method allows for the determination of the successes and failures of the mission depending on a number of input parameters. The results of the simulation must be averaged over several iterations to have a representative (median) outcome.

For all the simulations presented in this paper (with exception of Fig. 5), the results are averaged over 100 interstellar journeys. We briefly remind the reader that HERITAGE takes the following iterative steps: first it creates the initial crew according to the parameters that are listed in Table 1 (which, unless stated otherwise, are the one used in this paper). The code checks every year for accidental and natural deaths, then checks for every crew member that she/he is within the procreation window. If so, HERITAGE randomly associates two crew members of different sexes and evaluates if they can have a child. Infertility, pregnancy chances, inbreeding limitations and other parameters (such as the fact that the female crew member is not already pregnant) are verified before a successful mating. A new crew member is created in the vessel and the loop ends after surveying the whole female community that is within the procreation window. The code saves all the data onboard and starts a new year, until the completion of the interstellar mission. In this paper, the code parameters are the same as the ones used in the first paper of the series, except for the duration of the travel that corresponds to the necessary time to reach Proxima Centauri b at a velocity of 200 km.s$^{-1}$. We also set the date at which a plague-like catastrophic event happens as year 2500 after launch in order to check whether a catastrophe can lead the population to extinction. HERITAGE is entirely described in [14] and we advise the reader to refer to it for more details.

### 2.1 Using fixed social engineering principles

The anthropologist John Moore fixed a number of social engineering principles for the perpetuation of a multi-generational crew in a closed habitat. Namely, the starting crew should be young, childless married couples, allowing the crew to better adapt to their new environment before starting the reproduction. The second social engineering principle is to postpone parenthood until late in the female reproductive periods so that genetic variation is better maintained. The spacecraft would be then populated with smaller sibships and the age-sex distribution would be echeloned, reducing the number of non-reproductive young and old people, therefore stabilizing the social network. In the first paper of the series [14], we used different population estimations to model a 200-year journey. Namely, we considered a Moore population of 150 space settlers [15] and a Smith population of 14 000 humans [16]. In both cases the crew members were equally partitioned between women and men and all hand-picked to avoid initial consanguinity.

Running HERITAGE for such crews leads to a catastrophic outcome where all the crew disappears along the journey towards Proxima Centauri b, see Fig. 1. A very small population of 150 people dies after a tenth of the journey has been accomplished while a larger (14,000 humans) crew dies around the 1300th year (data averaged over 100 interstellar trips). The x-axis has been cut when the last human onboard disappeared. In this case[1], the decrease of population is due to the fixed number of children authorized by the social engineering principles (see Sect. 2.2). The main problem, identified in the first paper of this series, is that a too larger number of permitted children would lead to overpopulation. The complication arises from the severity of the rules that do not account for the actual conditions within the spacecraft at a given time. As it was concluded in [14], the fixed social engineering principles recommended by Moore are not adapted for long interstellar trips.

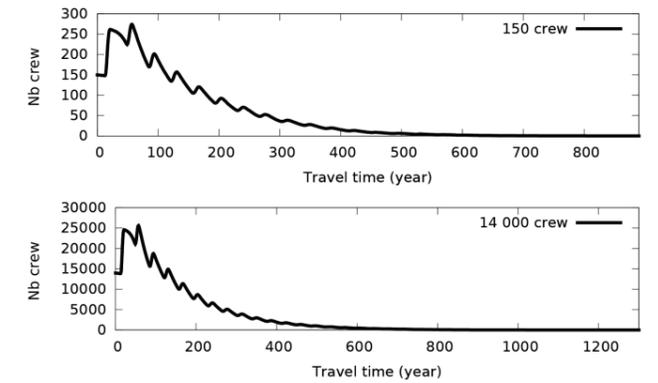

**Fig. 1** Crew population for a 6300-year trip where a fixed birth control is applied: an average number of 2 children is authorized, accounting for a standard deviation of 0.5. Top: a Moore-like population; bottom: a Smith-like population.

### 2.2 Using adaptive social engineering principles

We have shown in the previous section that the social rules must be adaptive, i.e., they must evolve with time and take into account the real situation onboard the vessel. To do so, the HERITAGE code was modified to account for adaptive social engineering principles: the average number of children per woman N (and the related standard deviation $\sigma$) is no longer constant but in evolution. Every year the code calculates the number of living people in the vessel and compares this number to a security threshold. This security threshold is set at the discretion of the user and represents the amount of people the vessel can support without suffering from overpopulation or scarcity of food/supplies. In this paper we will use a value of 90% of the colony ship capacity in order to have a security margin in the case of an accident such as the destruction of a given vessel sector, hence reducing the habitable space or supplies onboard.

If the amount of people inside the vessel is lower than the threshold, the code allows for a smooth increase of the population by allowing each woman to have an average of 3 children (with a standard deviation of 1). When the threshold is reached, HERITAGE impedes the couples' ability to procreate but allows women that were already pregnant to give birth even if the total number of crew members becomes marginally higher than the threshold. This is a moral procedure to prevent social tensions in the vessel due to imposed abortion. The impossibility to produce offspring is maintained until natural deaths happen, which cause the crew population number to be lower than the security threshold. Procreation can start again from this point.

In the orange curve of Fig.2 we see that the space crew does not survive much longer than a Smith-like population, despite the adaptive social engineering principles. Even if the initial crew is comprised of only 150 citizens, the narrow period allowed for procreation (between 35 and 40 years old) established by Moore is too restrictive. The age-sex distribution is

### TABLE 1  Input parameters of the simulation

| Parameter | Value | Units |
| --- | --- | --- |
| Number of space voyages to simulate | 100 | (integer) |
| Duration of the interstellar travel | 6300 | (years) |
| Colony ship capacity | 500 | (humans) |
| Number of initial women | 75 | (humans) |
| Number of initial men | 75 | (humans) |
| Age of the initial women | 20/1 | (years) |
| Age of the initial men | 20/1 | (years) |
| Women infertility | 0.10 | (fraction) |
| Men infertility | 0.15 | (fraction) |
| Number of child per woman | 2/0.5 | (humans) |
| Twinning rate | 0.015 | (fraction) |
| Life expectancy for women | 85/15 | (years) |
| Life expectancy for men | 79/15 | (years) |
| Mean age of menopause | 45 | (years) |
| Start of permitted procreation | 35 | (years) |
| End of permitted procreation | 40 | (years) |
| Chances of pregnancy after intercourse | 0.75 | (fraction per year) |
| Initial consanguinity | 0 | (fraction) |
| Allowed consanguinity | 1 | (fraction) |
| Life reduction due to consanguinity | 0.5 | (fraction) |
| Chaotic element of any human expedition | 0.001 | (fraction) |
| Possibility of a catastrophic event | 1 | (boolean) |
| Year at which the disaster will happen | 2500 | (year; 0 = random) |
| Fraction of the crew affected by the catastrophe | 0.30 | (fraction) |

**Note** The $\mu/\sigma$ values shown for certain parameters indicate that the code needs a mean ($\mu$) and a standard deviation value ($\sigma$) to sample a number from of a normal (Gaussian) distribution.

---

[1] Since the possible parameter space to explore for all possible fixed social principles is huge, we restrict ourselves to criteria based on the rules suggested by Moore.

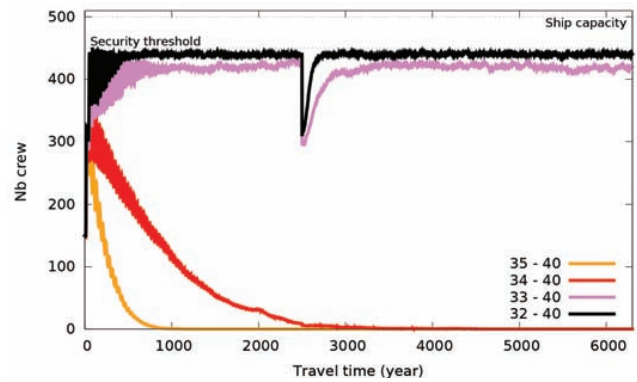

**Fig. 2** Crew population for a 6300-year trip where adaptive social engineering principles are accounted for. Different results, based on the period allowed for procreation, are shown. In orange is a simulation with procreation permitted between 35 and 40 years; in red the period is 34-40, in violet 33-40, and in black 32-40. The initial crew is composed of 75 women and 75 men.

echeloned during the first centuries and it is easy to find breeding partners within a diverse pool of crew members, but the dispersion allowed by the Gaussian distribution of children per woman ultimately averages the age-sex dispersion after several centuries. The amount of adults that are within the allowed procreation window becomes smaller and the pool for permitted reproduction decreases, leading the vessel towards slow extinction. This is the reason why the orange curve in Fig. 2 shows a spike followed by a collapse. It is thus mandatory to increase the allowed procreation window. The impact of the parameter phase space is shown in the same figure: in orange is a simulation with procreation permitted between 35 and 40 years, in red the period is 34-40, in violet 33-40, and in black 32-40. As shown in Fig. 2, a ship can have a stable population at the level of the security threshold if the period for procreation is equal to or larger than 33-40 years old. One can also see the impact of this parameter on the crew recovery after a cata-strophic event happening at year 2500. For larger procreation windows, the recovery time is shorter.

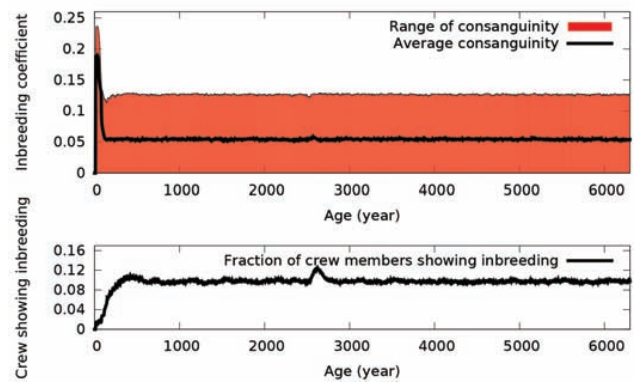

**Fig. 3** Inbreeding within the crew for a 6300-years trip toward Proxima Centauri b when adaptive social engineering principles are accounted for but without a control on authorized levels of consanguinity (see Fig. 2). Procreation is permitted between 32 and 40 years. Top: inbreeding coefficient as a function of time. The range of consanguinity (maximum- minimum) is shown in red and the average consanguinity factor $F$ per crew member is shown using the solid black line. The mean is measured from only those who show a non-zero coefficient. The dotted line at 0.05 represents the limit where deleterious effects onset. Bottom: fraction of crew members showing a non-zero consanguinity.

We thus saw that using adaptive social engineering principles, it is feasible to have a successful manned mission travelling for 6300 years towards Proxima Centauri b. However, the question of inbreeding was not yet discussed. In Fig. 3 below we show the inbreeding coefficient aboard the vessel as a function of time. The range of consanguinity (maximum-minimum) is shown in red and the average consanguinity factor $F$ per crew member is shown using the solid black line. The mean is measured from only those who show a non-zero coefficient and it should be compared to the inbreeding coefficients presented in Tab. 2. We find that, for an uncontrolled population, the average consanguinity factor per crew member lies between 6% and 6.5%, which corresponds to breeding between first cousins, half-uncle/niece or half-aunt/nephew (the procreation window preventing great-grandfather/great-granddaughter or great-grandmother/great-grandson mating). It is slightly larger than 5% – the limit where deleterious effects onset [17].

We observe an initial peak of high consanguinity (~ 18% on average) that happens during the first centuries. This corresponds to the first generations of space settlers, whose population number is relatively small and where random brother/sister mating can occur more often than when the onboard population reaches several hundreds of people. The high 18% averaged consanguinity factor quickly decreases when the population is big enough so that there are more chances to randomly mate between unrelated or distant-related pairs rather than brothers and sisters, leading to a stable population level.

The lower graph in Fig. 3 shows the fraction of crew members showing a non-zero consanguinity and we find that about 10% of the crew has signs of inbreeding. A 13% peak of inbreeding follows the restoration period consecutive to the catastrophic event but the remaining curve is plateauing at ~ 10%. While this is already a great success to have a 90% final crew still perfectly healthy after a 6300-year multi-generational journey, one may want to design a mission where the human genetic heritage is perfectly safe to ensure humankinds' survival. In conclusion, if the inbreeding coefficient does not reach highly dangerous levels, it remains questionable to have a genetically unhealthy crew landing on an extra-solar planet. For a purely genetic safety purpose, we will then restrict inbreeding within crew members in the following section.

### 2.3 Effects of inbreeding restrictions on the population

The necessity to restrict inbreeding is a conservative security condition to ensure a genetically healthy crew. Using HERITAGE, we explored the impact of controlled consanguinity within the crew and present the results in Fig. 4. Results are averaged over 100 space travel for each set of parameters. It appears that a spaceship where inbreeding is tolerated up to 10% is able to reach destination without any trouble. The averaged population is almost at the security threshold. However, when inbreeding is restricted to a value of 5% at maximum, the averaged population is lower (slightly larger than 400 space settlers). This effect is even more visible when inbreeding is strictly prohibited: the average population within the ship is about 320 people at the end of the journey. This is due to the fact that not all realizations of the journey are successful. A fraction of the simulations ended due to the inability of the crew to reproduce when following the inbreeding restrictions.

We illustrate the fact that not all interstellar travel reach destination in Fig. 5. We present 10 single-journey simulations randomly extracted from Fig. 4 (but the median of them being representative of the larger sample of 100). We see that there are two modes: low or negative early growth leading to complete failure after ~ 500 years, or steady initial growth leading to success. The random nature of births, deaths and mating, coupled to a strict rule of zero consanguinity allowed within the ship potentially drives the vessel to a doomed fate since, in several cases the pool for mating is not diverse enough. Three out of ten simulations failed before the first millennium of space mission. After 1000 years of interstellar travel, we observe no failures at all. Since we chose the disaster to occur relatively late in the mission, it has little effect on the total population and never leads to a mission failure. By 2500 years, if any of the crew have survived then their population has invariably reached a level so high that reducing them by one-third still leaves more than sufficient breeding pairs for a complete recovery (see Sect. 2.4). This suggests that the crew could potentially survive multiple such disasters, provided their frequency was sufficiently low. Of course, if they occurred too frequently or too early in the mission (when the population was still low), or caused more fatalities than we have stipulated, then they could potentially indeed prove fatal, but we do not explore this in detail here. All successful missions actually have a population level very close to the security threshold but the average of the ten expeditions results in an crew population lower than the security threshold such as seen in Fig. 4 (30% of failure leads to 0.3 × Security Threshold ≈ 300). The reader must keep in mind that the simulation can only have two outcomes: success or failure. Such interstellar mission should be launched with a 100% success rate and the Monte Carlo method allows us to determine which initial conditions are needed to achieve this goal.

**TABLE 2** Inbreeding coefficients $F$

| Relationship | $F$ |
| --- | --- |
| Identical twins | 100% |
| Self fertilization | 50% |
| Brother/sister | 25% |
| Father/daughter or mother/son | 25% |
| Grandfather/granddaughter or grandmother/grandson | 12.5% |
| Half-brother/half-sister | 12.5% |
| Uncle/niece or aunt/nephew | 12.5% |
| Great-grandfather/great-granddaughter or great-grandmother/great-grandson | 6.25% |
| Half-uncle/niece or half-aunt/nephew | 6.25% |
| First cousins | 6.25% |
| First cousins once removed or half-first cousins | 3.125% |
| Second cousins or first cousins twice removed | 1.5625% |
| Second cousins once removed or half-second cousins | 0.78125% |
| Third cousins or second cousins twice removed | 0.390625% |
| Third cousins once removed or half-third cousins | 0.195% |
| **Note** Values of the inbreeding coefficients $F$ for consanguineous matings (one generation, no previous in-breeding). | |

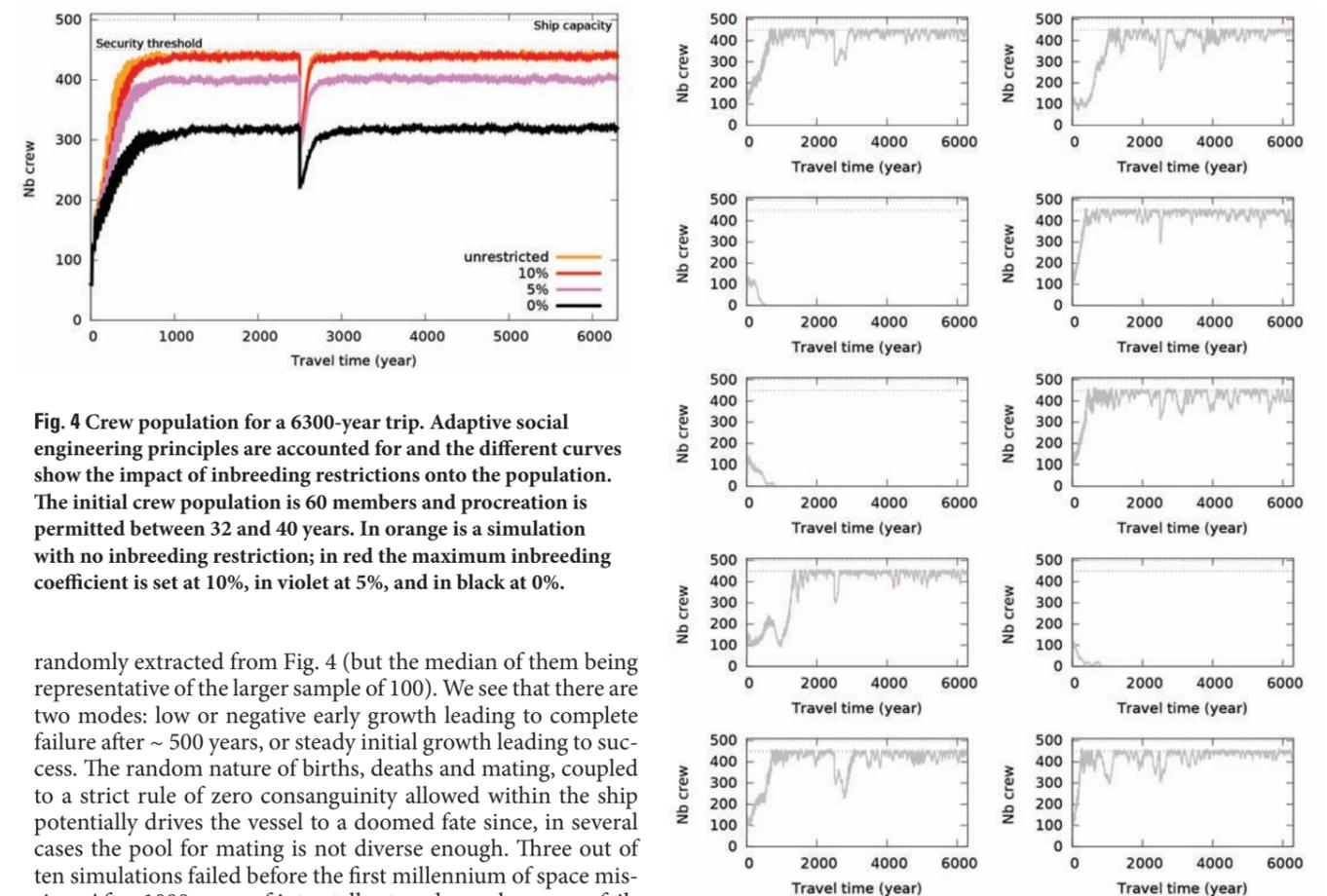

**Fig. 4** Crew population for a 6300-year trip. Adaptive social engineering principles are accounted for and the different curves show the impact of inbreeding restrictions onto the population. The initial crew population is 60 members and procreation is permitted between 32 and 40 years. In orange is a simulation with no inbreeding restriction; in red the maximum inbreeding coefficient is set at 10%, in violet at 5%, and in black at 0%.

**Fig. 5** Ten single-journey realizations extracted from Fig. 4. The allowed consanguinity was set to 0%, the initial crew population is 60 and procreation was permitted between 32 and 40 years.

### 2.4 Estimating the success/failure rate

Due to the randomization of events within its Monte Carlo architecture, HERITAGE is able to estimate if a mission is destined to succeed or fail if the simulations are looped over several dozen attempts. In the following, we will estimate the success rate of a multi-generational space ship with different initial crew populations. We fix the ratio of women and men to parity and we allow procreation to happen between 32 and 40 years. While it is debatable whether a crew with zero consanguinity is mandatory or if a few crew members can show small (i.e. < 5%) inbreeding coefficient, we restrict inbreeding to a null quantity for conservative reasons. By doing so we can focus on the most constrained simulations. We looped HERITAGE one hundred times for each simulation in order to have statistically significant error bars.

The results of our investigation are presented in Fig. 6. The standard deviation to the mean of the success rate is presented at 3-$\sigma$ (i.e., there is a 99.7% probability that the estimated success rate is certain). We can see that, for less than 32 initial crew members in the vessel, the simulation gives a chance of success that can reach 0% as inbreeding cannot be prevented in such small communities. With larger initial crews, the chances of reaching the ship destination with a healthy crew increases. The slope appears to be linear, with small variations due to the statistical approach.

A mission has about a 50 ± 15% chance of being successful if the initial crew is composed of 25 women and 25 men (25 breeding pairs). It has been observed in laboratories that the genetic diversity of a colony of animals (but this could be applied to humans as well) composed of 25 pairs can be sustained practically infinitely by careful pairing, especially if spontaneous mutations are taken into account [18]. In comparison to random mating using at least 25 breeding pairs per generation, a consistent order rotational breeding scheme allows the creation of an outbred stock (a colony within which there is some genetic variation and which has been closed for at least four generations) with half these numbers [19].

This is the reason why HERITAGE, based on random pairing, predicts that 0% of the missions carrying less than 14 initial breeding pairs can reach Proxima Centauri b. A rotational breeding scheme control would be questionable morally, so in case of populating a distant planet, a larger group is needed to provide for sexual preferences, fertility problems, sudden deaths, etc. Finally, we see from Fig. 6 that the success rate reaches a plateau of 94-98% chance of success when the initial crew population is equal to or exceeds 68 members. The code clearly indicates that unpredictable accidents may happen during the flight time that would ultimately wipe out a population even if the initial crew population is moderately large. In order to ensure a safe multi-generational space travel towards Proxima Centauri b, a minimum initial crew of 98 settlers is necessary (under the input parametrization presented in this paper).

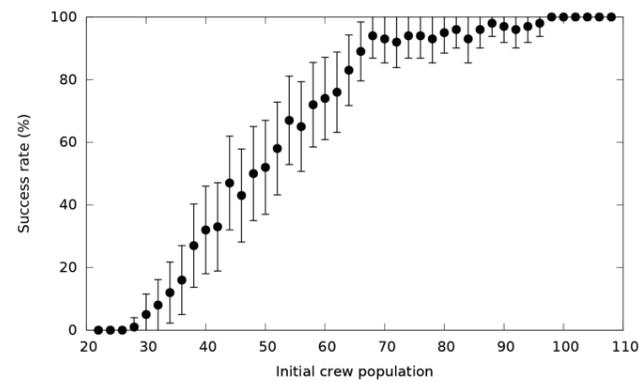

**Fig. 6** Success rate of a 6300-year mission towards Proxima Centauri b as a function of the initial crew population. The results and the standard 3-$\sigma$ significance of the reported proportions are estimated by averaging 100 simulations for each population number. The allowed consanguinity was set to 0% and procreation was permitted between 32 and 40 years.

## 3 DISCUSSIONS

We have seen throughout this paper that space travel to Proxima Centauri b is feasible under a number of given conditions. First, any multi-generational crew must follow social engineering principles in order to be able to survive centuries-long journeys. However, those rules must be adaptive, otherwise the population is slowly doomed to a fatal end as inbreeding or overcrowding occur. By means of adaptive social engineering principles, where the number of offspring is controlled yearly within the ship, it is possible to regulate the population level. A security threshold must be defined so that the population always stays under the maximal limit imposed by the size of the vessel itself. As expected, we found that restrictions on the consanguinity levels have a profound impact on the chances of success of the mission. It is important to maintain a genetically healthy crew, but this may drive the breeding scheme to ultra-conservative constraints. By restricting the offspring to being genetically pure, we found that 50±15% missions successfully reach their destination for restricted mating between 25 initial breeding pairs. If the initial crew is comprised of 34 initial breeding pairs, the chances to reach their destination with a completely genetically healthy crew rise up to 94-98%. An initial ship with 98 settlers (49 initial breeding pairs) ensures a mission with a 100% success rate at 3-$\sigma$ significance. We can then conclude that, under the parameters used for those simulations, a minimum crew of 98 settlers is needed for a 6300-year multi-generational space journey towards Proxima Centauri b.

### 3.1 Comparisons with previously published results

In this paper, we found a minimum crew that is less than what was advocated by Moore [15] and Smith [16]. In the former case the minimal initial crew population was estimated at about 150 - 180 people, and in the second case the numbers are larger, between 14000 and 44000 people. If the simulation by Moore relies on numerical methods, the mating scheme we use here is much more efficient since we try to minimize the effects of inbreeding by not assigning each female member to a unique male partner. The very narrow procreation window used by Moore naturally drove the initial population to larger numbers since the necessity for gene diversity was higher. By relaxing the procreation window and perfecting the mating scheme, the final population of initial crew member logically decreases. The case of Smith is quite different. First the 14000 - 44000 people estimation do not rely on Monte Carlo computations; they are found by multiplying statistics. The absence of repeated and averaged Monte Carlo simulations prevent a real determination of the initial crew needed for a multi-generational interstellar mission. On the other hand the great achievement of Smith is to maximize genetic diversity. The impact of mutation, migration, selection and drift is not included in HERITAGE. For this reason we emphasize that the minimum crew of 98 settlers we found is a lower limit under the conservative conditions specified in this paper. The initial crew will be smaller if we allow a small consanguinity within the crew members but it might get larger once the central issues of population genetics will be accounted for.

### 3.2 On the habitability of Proxima Centauri b

The question of the real potential habitability of Proxima Centauri b is to be investigated before any interstellar expedition. The fact that Proxima Centauri b closely orbits a red dwarf star might be problematic. During the pre-main sequence phase, the change of irradiation due to the star evolution may have had a strong impact on the exoplanet. The greenhouse gases (if any) could have been wiped out, leading to a planet subject to X-ray and extreme ultraviolet irradiation (and strong stellar winds). Any water molecules in the atmosphere or surface of Proxima Centauri b could have evaporated, resulting in a barren, Venus-like exoplanet [20, 21]. This scenario is balanced by the unknown true complex evolutionary history of Proxima Centauri b that cannot be probed yet. The fate of Proxima Centauri b strongly depends upon its initial water content (which is not constrained by planet formation models), the amount and composition of its initial gaseous envelope and its possible migration with respect to its host star. Varying one or several of those aspects can lead the planet to be safely habitable [20, 22, 23]. The presence of one or multiple exomoons in stable circular orbits around the exoplanet may affect the habitability of the target [24, 25] and remains to be characterized and observed for extra-solar systems. Atmospheric characterization will be possible in the future thanks to the European Extremely Large Telescope (E-ELT) that will enable high-resolution spectroscopy of the exoplanet atmosphere. Infrared interferometers and spatial missions with high resolution detectors will look for tracers and molecular signatures such as $H_2O$, $O_2$, and $CO_2$ to determine whether the surface of this exoplanet is habitable.

## 4 CONCLUSIONS AND FURTHER DEVELOPMENT

Hence, if Proxima Centauri b is really habitable and if future multi-generational spacecraft are travelling at a human-feasible speed of 0.067% of the speed of light, any vessel should embark at least 98 initial crew members (under the conservative conditions specified in this paper). To pursue those investigations, it is necessary to look at other fine details. In particular, the central issues of population genetics (effects of mutation, migration, selection and drift) have to be included in HERITAGE [16], together with better estimates of the process of reproductions based on clinical experimental results. Finally, linked with gene deterioration and health issues, the impact of cosmic rays has to be taken into account. While a frontal shield will be surely included in all vessel designs to prevent irradiation by high energy cosmic particles, the deterioration and natural wear of the shield may impact the population onboard. All those subjects will be investigated in forthcoming publications based on HERITAGE.

### Acknowledgement

The authors would like to thank Dr. Rhys Taylor (astrophysicist) for his careful reading of the manuscript and for his numerous comments that helped to improve this article.